\documentclass[aps,prb,twocolumn,superscriptaddress]{revtex4}

\usepackage{graphicx}
\usepackage{bm}

\begin{document}

\preprint{}

\title{Enhancement of Superconducting Transition Temperature Due to Antiferromagnetic Spin Fluctuations in Iron-pnictides LaFe(As$_{1-x}$P$_x$)(O$_{1-y}$F$_y$) : $^{31}$P-NMR Studies}

\author{H. Mukuda}\email[]{e-mail  address: mukuda@mp.es.osaka-u.ac.jp}
\author{F. Engetsu}
\author{K. Yamamoto}
\affiliation{Graduate School of Engineering Science, Osaka University, Osaka 560-8531, Japan}
\author{K. T. Lai}
\affiliation{Graduate school of Science, Osaka University, Osaka 560-0043, Japan}
\author{M. Yashima}
\author{Y. Kitaoka}
\affiliation{Graduate School of Engineering Science, Osaka University, Osaka 560-8531, Japan}
\author{\\A. Takemori}
\author{S. Miyasaka}
\author{S. Tajima}
\affiliation{Graduate school of Science, Osaka University, Osaka 560-0043, Japan}

\date{\today}

\begin{abstract}

Systematic $^{31}$P-NMR studies on LaFe(As$_{1-x}$P$_x$)(O$_{1-y}$F$_y$) with $y$=0.05 and 0.1 have revealed that the antiferromagnetic spin fluctuations (AFMSFs) at low energies are markedly enhanced around $x$=0.6 and 0.4, respectively, and as a result, $T_c$ exhibits respective peaks at 24 and 27 K against the P substitution for As. 
This result demonstrates that the AFMSFs are responsible for the  increase in $T_c$ for LaFe(As$_{1-x}$P$_x$)(O$_{1-y}$F$_y$) as a primary mediator of the Cooper pairing.  
From a systematic comparison of AFMSFs with a series of (La$_{1-z}$Y$_z$)FeAsO$_{\delta}$ compounds in which $T_c$ reaches 50 K for $z$=0.95, we remark that a moderate development of AFMSFs causes $T_c$ to increase up to 50 K under the condition that the local lattice parameters of the FeAs tetrahedron approach those of the regular tetrahedron.  We propose that $T_c$ of Fe-pnictides exceeding 50 K is maximized under an intimate collaboration of the AFMSFs and other factors originating from the optimization of the local structure. 

\end{abstract}


\maketitle

\section{Introduction}

The iron-based oxypnictide LaFeAsO, which is an antiferromagnet with an orthorhombic structure, becomes a superconductor at transition temperature $T_c$=26 K for LaFeAsO$_{1-y}$F$_y$(La1111) by the substitution of O$^{2-}$ with F$^-$ when $y$=0.1\cite{Kamihara2008,Luetkens}. 
Since its discovery, the role of antiferromagnetic spin fluctuations (AFMSFs) is believed to be indispensable for the onset of superconductivity (SC). 
On the other hand, it was reported that $T_{c}$ reaches a maximum of 55 K for the Sm1111 compound\cite{Ren1,Ren2}, in which the  FeAs$_{4}$ block forms into a nearly regular tetrahedral structure~\cite{C.H.Lee}. In this structure, the optimal values for the lattice parameters, which enhance $T_c$, are the As-Fe-As bonding angle $\alpha$=109.5$^\circ$~\cite{C.H.Lee}, the height of the pnictogen $h_{Pn}\sim$1.38\AA~from the Fe plane~\cite{Mizuguchi}, and the $a$-axis length $a\sim$3.9\AA~\cite{Ren2,Miyazawa1}. This regular tetrahedral structure is expected to yield a multiplicity of the Fermi surface topology, multiple excitations that are relevant to the $d$-orbital degeneracy, and fluctuations of $d$ orbital and spin degrees of freedom. 

\begin{figure}[htbp]
\centering
\includegraphics[width=8cm]{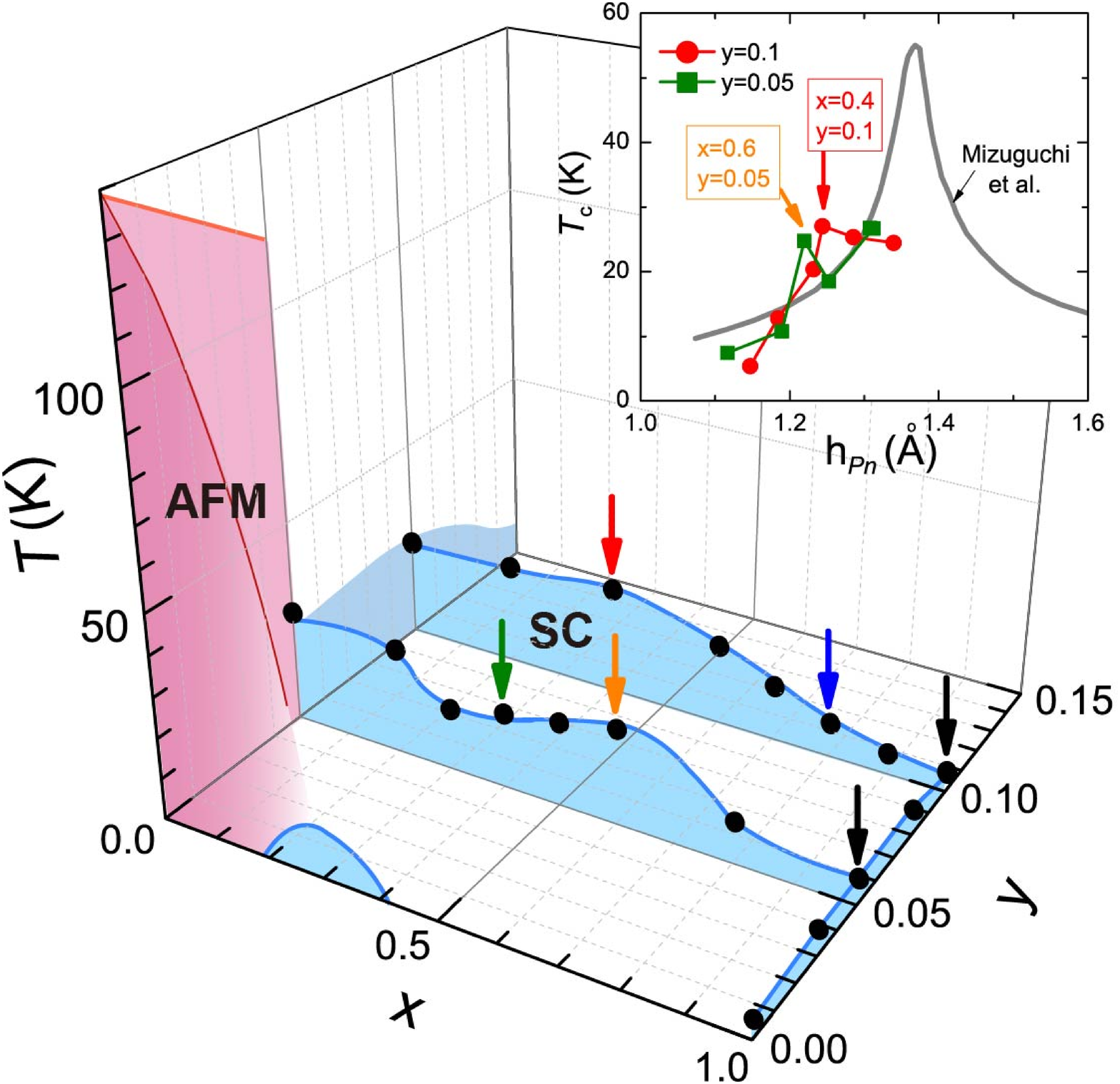}
\caption[]{(Color online) 
Phase diagram of LaFe(As$_{1-x}$P$_x$)(O$_{1-y}$F$_y$). The $T_c$ and $T_N$ values come from the previous works on  LaFe(As$_{1-x}$P$_x$)(O$_{0.9}$F$_{0.1}$)\cite{Saijo,Miyasaka,Lai}, LaFe(As$_{1-x}$P$_x$)O\cite{Wang}, LaFeP(O$_{1-y}$F$_{y}$)\cite{SSuzuki}, and LaFeAs (O$_{1-y}$F$_{y}$)\cite{Kamihara2008,Luetkens}. The arrows indicate the content for the samples used here. The inset shows $T_c$ vs $h_{Pn}$ for LaFe(As$_{1-x}$P$_x$)(O$_{1-y}$F$_{y}$) plotted on a universal relation for many Fe-pnictides reported by Mizuguchi {\it et al.}\cite{Mizuguchi}.    
}
\label{PhaseDiagram}
\end{figure}

In order to shed further light on an interplay between AFMSFs and fluctuations originating from the local degrees of freedom, we present normal-state and SC characteristics probed by a $^{31}$P-NMR for a series of LaFe(As$_{1-x}$P$_x$)(O$_{1-y}$F$_{y}$) compounds with $y$=0.1 and 0.05. The isostructural compound  LaFePO exhibits an SC transition at $T_c\sim$ 4 K without any substitution, however, a partial replacement of O$^{2-}$ with F$^-$ causes $T_c$ to increase to 7 K~\cite{Kamihara2006}. 
In LaFe(As$_{1-x}$P$_x$)(O$_{0.9}$F$_{0.1}$), which are all superconductive~\cite{Saijo,Miyasaka}, $T_c$ reaches a maximum of 27 K at $x$=0.4 as shown in Fig. \ref{PhaseDiagram}, even though the lattice parameters  are monotonously varied with $x$ and are apart from the optimum values of the FeAs$_{4}$ block~\cite{Saijo,Lai}. 
In this context, further systematic studies on these LaFe(As$_{1-x}$P$_x$)(O$_{1-y}$F$_{y}$) compounds provide us with an opportunity to identify a possible parameter for raising the $T_c$, apart from the optimization of the local structure of the Fe-based superconductors.
In fact, here we report that as a consequence of the development of the AFMSFs at low energies for compounds at $x$=0.4 and $y$=0.1, the $T_c$ increases to 27 K, which is higher than the $T_c$ for the original compound at $x$=0 and $y$=0.1. 
However, when AFMSFs are not visible, the $T_c$ at $x$=1.0 decreases to 5.4 K. 
Similar results have been obtained for the underdoped compounds at $y$=0.05. 
Present studies reveal that the AFMSFs are indispensable for raising the $T_c$ in LaFe(As$_{1-x}$P$_x$)(O$_{1-y}$F$_y$) compounds. 

\section{Experiment}

Polycrystalline samples of LaFe(As$_{1-x}$P$_x$)(O$_{1-y}$F$_y$) were synthesized by the solid-state reaction method, as described elsewhere~\cite{Saijo,Lai}. 
Powder X-ray diffraction measurements indicate that these samples are comprised of a single phase.  
Bulk $T_{\rm c}$s  were determined from an onset of SC diamagnetism in the susceptibility measurement. 
As shown in  Fig.~\ref{PhaseDiagram}, the $T_{\rm c}$ exhibits a maximum at $x$=0.4 for $y$=0.1\cite{Saijo,Miyasaka}, however, they exhibit a shallow minimum around $x$=0.3$\sim$0.4 and a local maximum at $x$=0.6 for $y$=0.05\cite{Lai}. 
$^{31}$P-NMR($I=1/2$) measurements on these compounds have been performed on coarse powder samples with a nominal content of $x$=0.4($T_c$=27 K), $x$=0.8($T_c$=8.8 K), and $x$=1.0($T_c$=5.4 K) for $y$=0.1, and $x$=0.4($T_c$=19 K), $x$=0.6($T_c$=24 K) and $x$=1.0($T_c$=6.7 K) for $y$=0.05, as indicated by the arrows in Fig.~\ref{PhaseDiagram}.  
The respective values of $a$-axis length, $h_{Pn}$, and $\alpha$ in LaFe(As$_{1-x}$P$_x$)(O$_{1-y}$F$_y$) monotonously vary from 4.002\AA, 1.24\AA, and 116.3$^\circ$ for $x$=0.4 to 3.951\AA, 1.15\AA, and 119.7$^\circ$ for $x$=1.0 when $y$=0.1, and from 4.011\AA, 1.25\AA, and 116.0$^\circ$ for $x$=0.4 to 3.959\AA, 1.12\AA, and 121.1$^\circ$ for $x$=1.0 when $y$=0.05\cite{Saijo,Lai}. 

\begin{figure}[htbp]
\centering
\includegraphics[width=7cm]{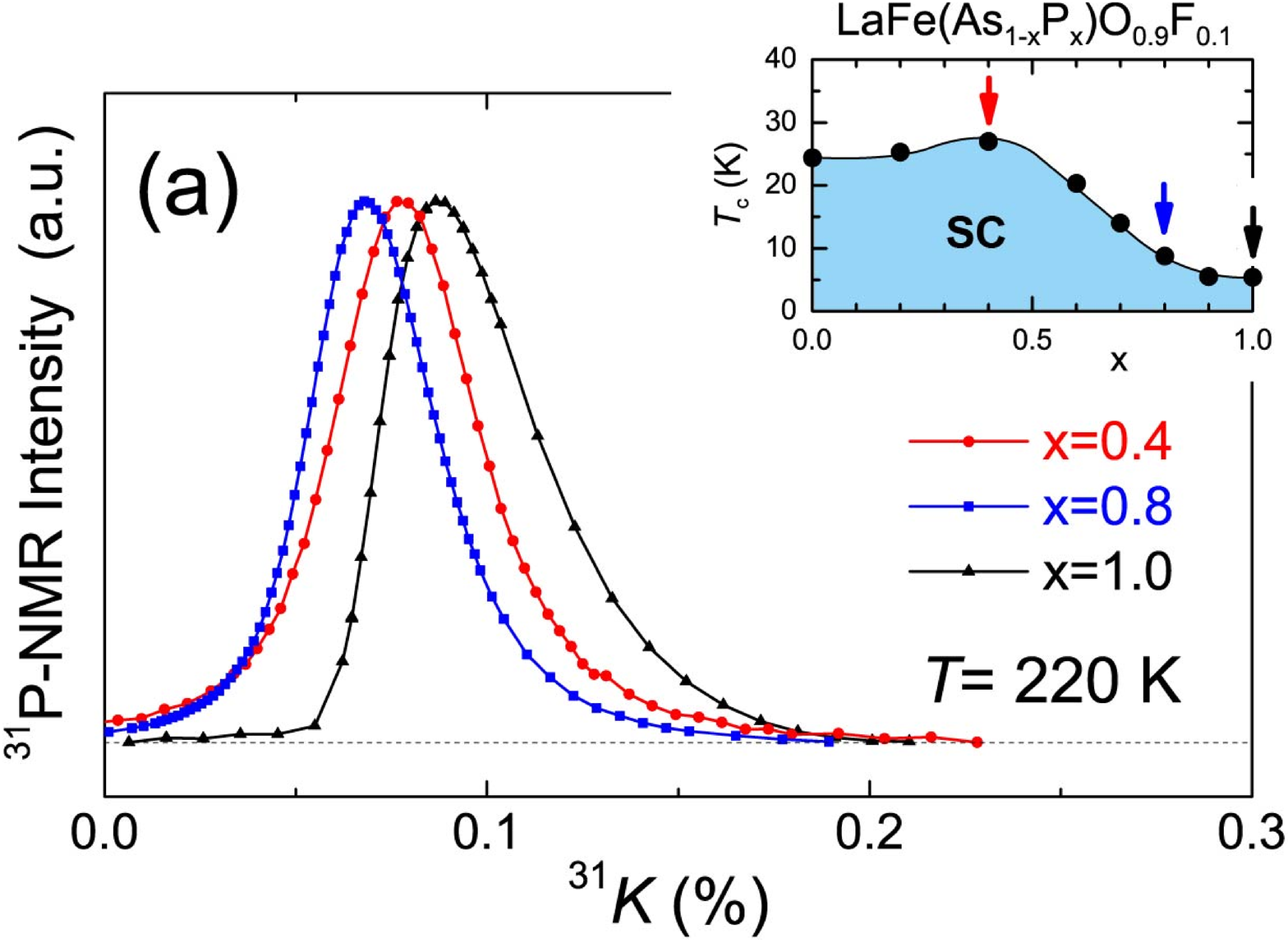}
\includegraphics[width=7cm]{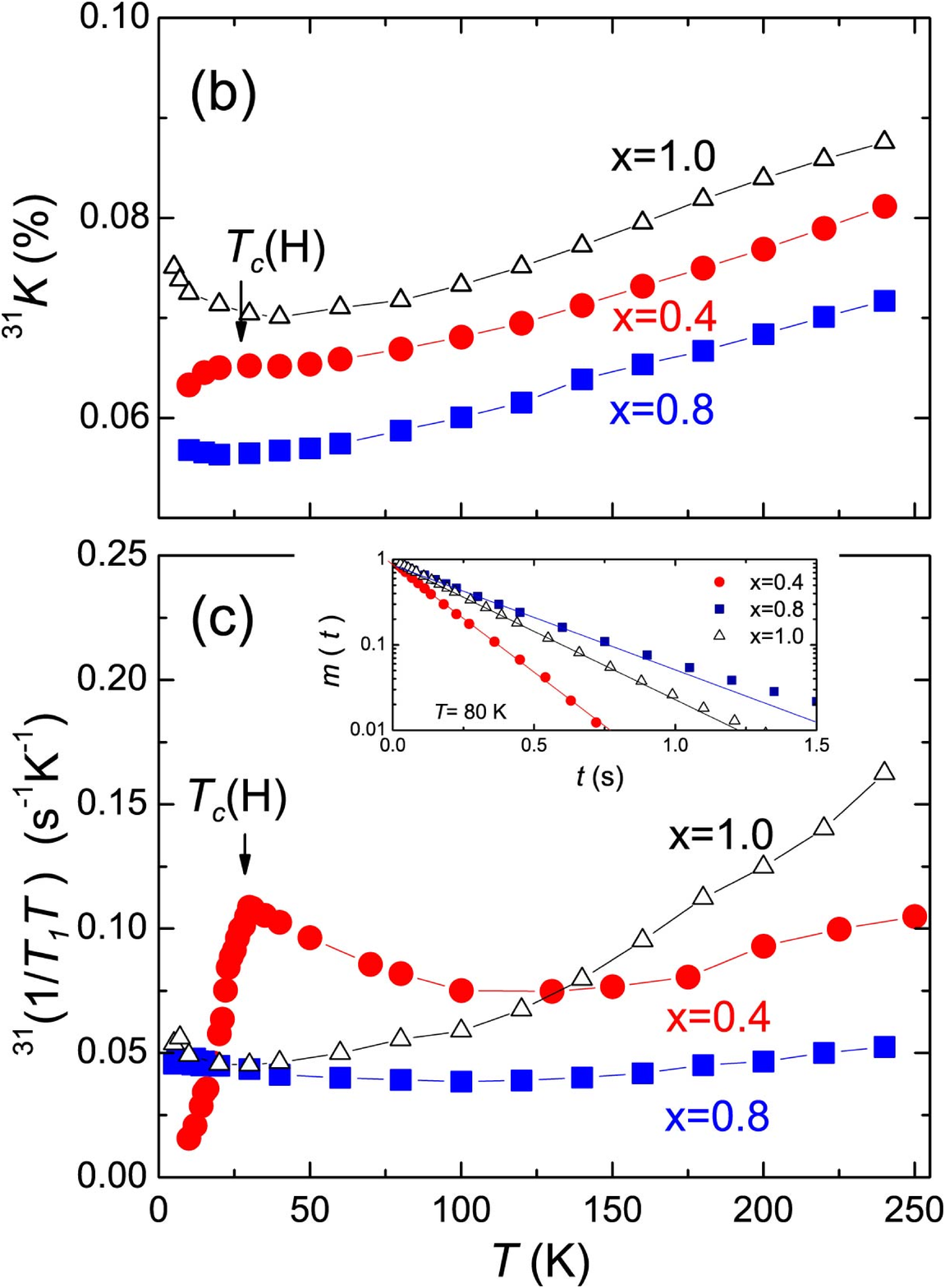}
\caption[]{(Color online) (a) $^{31}$P-NMR spectra at $T$=220 K, and $T$ dependence of the (b) Knight shift ($^{31}K$) and (c) $^{31}(1/T_1T)$ for LaFe(As$_{1-x}$P$_x$)(O$_{0.9}$F$_{0.1}$). The arrows in the inset of (a) indicate the samples used in this experiment. $T_c(H)$ indicates $T_c$ under the field $H\sim$11.95 T.  The inset in (c) shows the typical  recovery curves of the nuclear magnetization to determine $1/T_1$. 
} 
\label{NMR_F01}
\end{figure}

The $^{31}$P-NMR Knight shift $^{31}K$ was measured under a magnetic field of $\sim$11.95 T, which was calibrated by a resonance field of $^{31}$P in H$_3$PO$_4$. 
The nuclear-spin-lattice-relaxation rate $^{31}$($1/T_1$) of $^{31}$P-NMR was obtained by fitting a recovery curve of $^{31}$P nuclear magnetization to a single exponential function $m(t)\equiv (M_0-M(t))/M_0=\exp \left(-t/T_1\right)$. 
Here, $M_0$ and $M(t)$ are the nuclear magnetizations for a thermal equilibrium condition and at time $t$ after the saturation pulse, respectively. 
Note, however, that $m(t)$ in some compounds includes two components in $1/T_1$, as shown in the inset of Figs. \ref{NMR_F01}(c) and \ref{NMR_F005}(c), due to some inevitable inhomogeneity of the electronic states in association with the chemical substitution of P for As. 
Here, since the fraction of the short component of $1/T_1$ was predominantly larger than the long one, $1/T_1$ was determined by the short component. 

\section{Results}

\subsection{LaFe(As$_{1-x}$P$_x$)(O$_{0.9}$F$_{0.1}$)}

Figure \ref{NMR_F01}(a) shows the $^{31}$P-NMR spectra at $T$= 220 K for $x$=0.4, 0.8, and 1.0 of LaFe(As$_{1-x}$P$_x$)(O$_{0.9}$F$_{0.1}$). 
The full-width at half maximum (FWHM) of the $^{31}$P-NMR spectra is quite narrow, for example, $\sim$90($\sim$79) kHz at $x$=0.4 ($x$=0.8) at the resonance frequency  $\sim$206 MHz. 
Figures~\ref{NMR_F01}(b) and \ref{NMR_F01}(c) show the $T$ dependence of Knight shift ($^{31}K$) and $^{31}(1/T_1T)$, respectively, for $x$=0.4, 0.8, and 1.0 of LaFe(As$_{1-x}$P$_x$)(O$_{0.9}$F$_{0.1}$). 
Both the $^{31}K$ and $^{31}(1/T_1T)$ gradually decrease upon cooling at high temperatures, in contrast to that at low temperatures where the $T$ dependence of $^{31}(1/T_1T)$ strongly depends on the $x$. 

\begin{figure}[htbp]
\centering
\includegraphics[width=8cm]{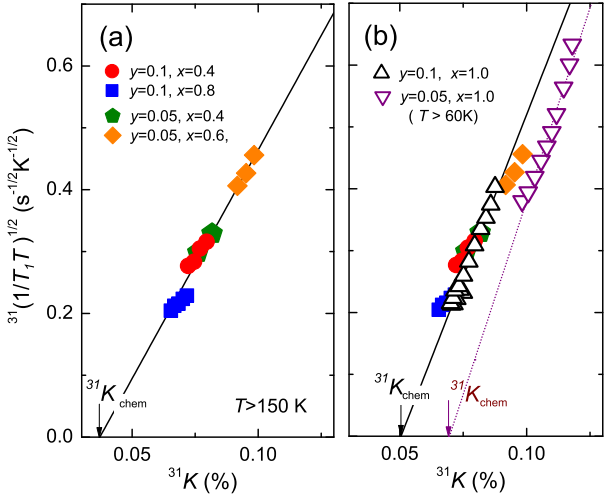}
\caption[]{(Color online) Plot of $^{31}(1/T_1T)^{1/2}$ vs. $^{31}K$ with an  implicit parameter of $T$. (a) For  $x$=0.4 and 0.8 at $y$=0.1  and  $x$=0.4 and 0.6 at $y$=0.05,  the $T$-independent $^{31}K_{chem}$ was evaluated to be 0.037\% using the data of $T>$ 150 K since the AFMSFs develop below 100 K. (b) For $x$=1.0, $^{31}K_{chem}$ was evaluated to be  0.05\% in the $T$ range of $T>$ 60 K for $y$=0.05 and 0.07\% in the whole $T$ range for $y$=0.1. 
} 
\label{Kchem}
\end{figure}
\begin{figure}[htbp]
\centering
\includegraphics[width=8cm]{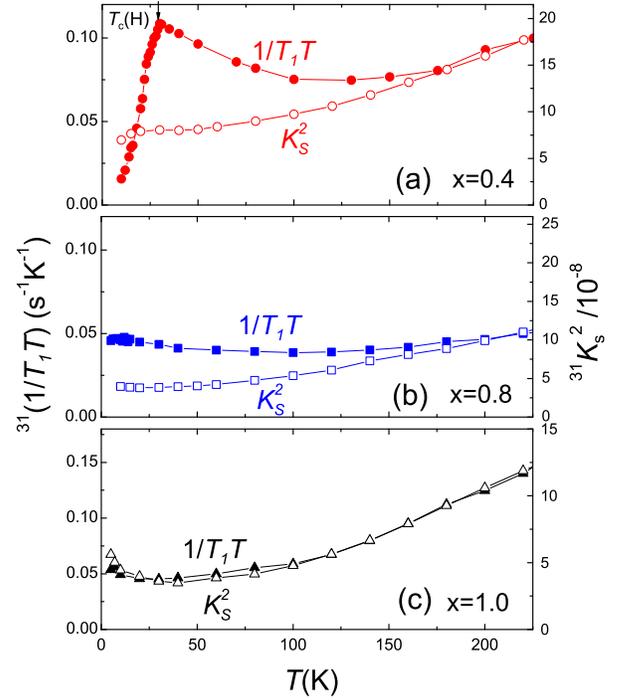}
\caption[]{(Color online) $T$ dependence of $^{31}K_s^2$(=$(^{31}K-^{31}K_{chem})^2)$ and $^{31}(1/T_1T)$ for (a)$x$=0.4, (b) $x$=0.8, and (c) $x$=1.0 of LaFe(As$_{1-x}$P$_x$)(O$_{0.9}$F$_{0.1}$). 
} 
\label{NMR_F010_2}
\end{figure}

The Knight shift comprises the $T$-dependent spin shift  $^{31}K_s(T)$  and the $T$-independent chemical shift $^{31}K_{chem}$. 
The former $^{31}K_s$($T$) is given using the static spin susceptibility $\chi({\bm q}=0)$ by 
\[
K_s(T) \propto A_{q=0}\ \chi({\bm q}=0) \propto A_{q=0} N(E_F),
\]
where $A_{q=0}$ is the hyperfine-coupling constant for the $q$=0 wave number and $N(E_F)$ is the density of states (DOS) at the Fermi level ($E_F$). 
In the nonmagnetic compounds, $^{31}K_s$ is proportional to $^{31}(1/T_1T)^{0.5}$ since Korringa's relation $^{31}(1/T_1T)\propto N(E_F)^2$ holds. 
As shown in Fig. \ref{Kchem}, the plot of $^{31}(1/T_1T)^{0.5}$ and $^{31}K$ enables us to evaluate $^{31}K_{chem}$  to be $\sim$0.05\% for $x$=1.0 using the data in whole $T$ range and $\sim$0.037\% for $x$=0.4 and 0.8 using the data at high temperatures ($T>$150 K), where the contribution of AFMSFs in $1/T_1T$ is negligible. 
The $^{31}K_s$($T$) that is evaluated from the relation of $^{31}K-^{31}K_{chem}$ decreases as the temperature lowers, as is observed for most electron-doped compounds~\cite{Grafe,Terasaki,Ning,Imai}. 
It is due to the narrow peak of the DOS being located below the $E_F$, which is the characteristic band structure for  electron-doped systems \cite{Ikeda}.   
In general, $1/T_1T$ can be expressed as 
\[
\frac{1}{T_1T}\propto \lim_{\omega_0 \rightarrow 0} \sum_{\bm q} |A_{\bm q}|^2 \frac{\chi''({\bm q},\omega_0)}{\omega_0},
\] 
where $A_{\bm q}$ is the ${\bm q}$-dependent hyperfine-coupling constant, $\chi({\bm q},\omega)$ is the dynamical spin susceptibility, and $\omega_0$ is the NMR frequency.   
Note that $1/T_1T$ is dominated by spin fluctuations at the low-energy limit since the NMR frequency $\omega_0$ is as low as a radio frequency. 
Figures \ref{NMR_F010_2}(a), \ref{NMR_F010_2}(b), \ref{NMR_F010_2}(c) show the $T$ dependence of $^{31}(1/T_1T)$ and $^{31}K_s^2$ for $x$=0.4, $x$=0.8, and $x$=1.0, respectively. 
$^{31}K_s(T)$, which is proportional to $\chi({\bm q}=0)$, decreases upon cooling, whereas $^{31}(1/T_1T)$ at $x$=0.4 increases  up to $T_c(H)$ upon cooling below 100 K, indicating that the development of AFMSFs occurred at a finite $Q$ wave vector presumably around ($\pm\pi$, 0) and (0, $\pm\pi$) \cite{Kuroki2}. 
By contrast, such an increase of $^{31}(1/T_1T)$ at low temperature is gradually suppressed at $x$=0.8  and considerably suppressed at $x$=1.0, where the decrease of $^{31}(1/T_1T)$ upon cooling is almost the same as that of $^{31}K_s^2$. 
The results demonstrate that strong AFMSFs at $x$=0.4  that exhibit  higher $T_c$  gradually decrease toward $x$=1.0  with  lower $T_c$.

\begin{figure}[htbp]
\centering
\includegraphics[width=7cm]{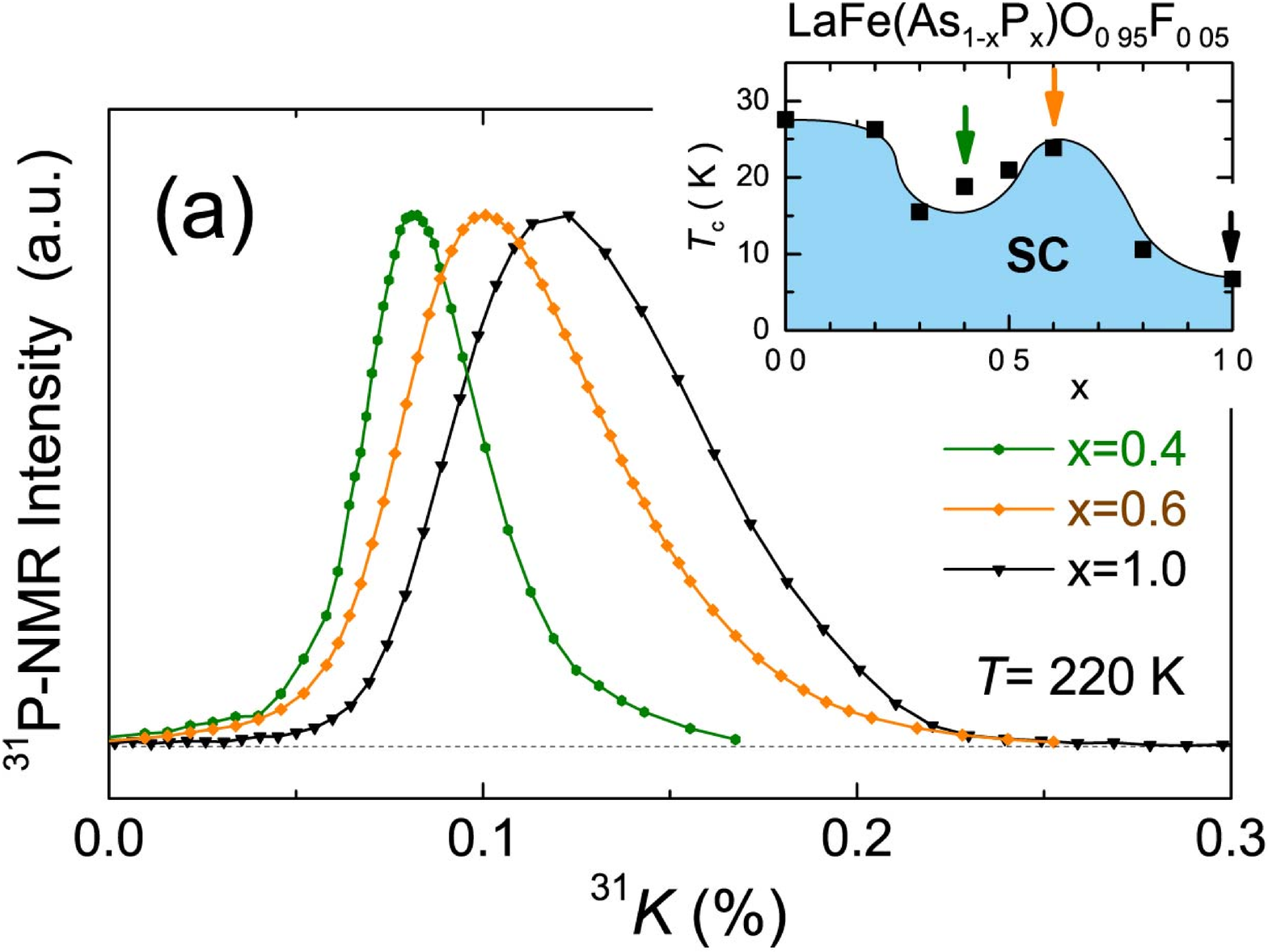}
\includegraphics[width=7cm]{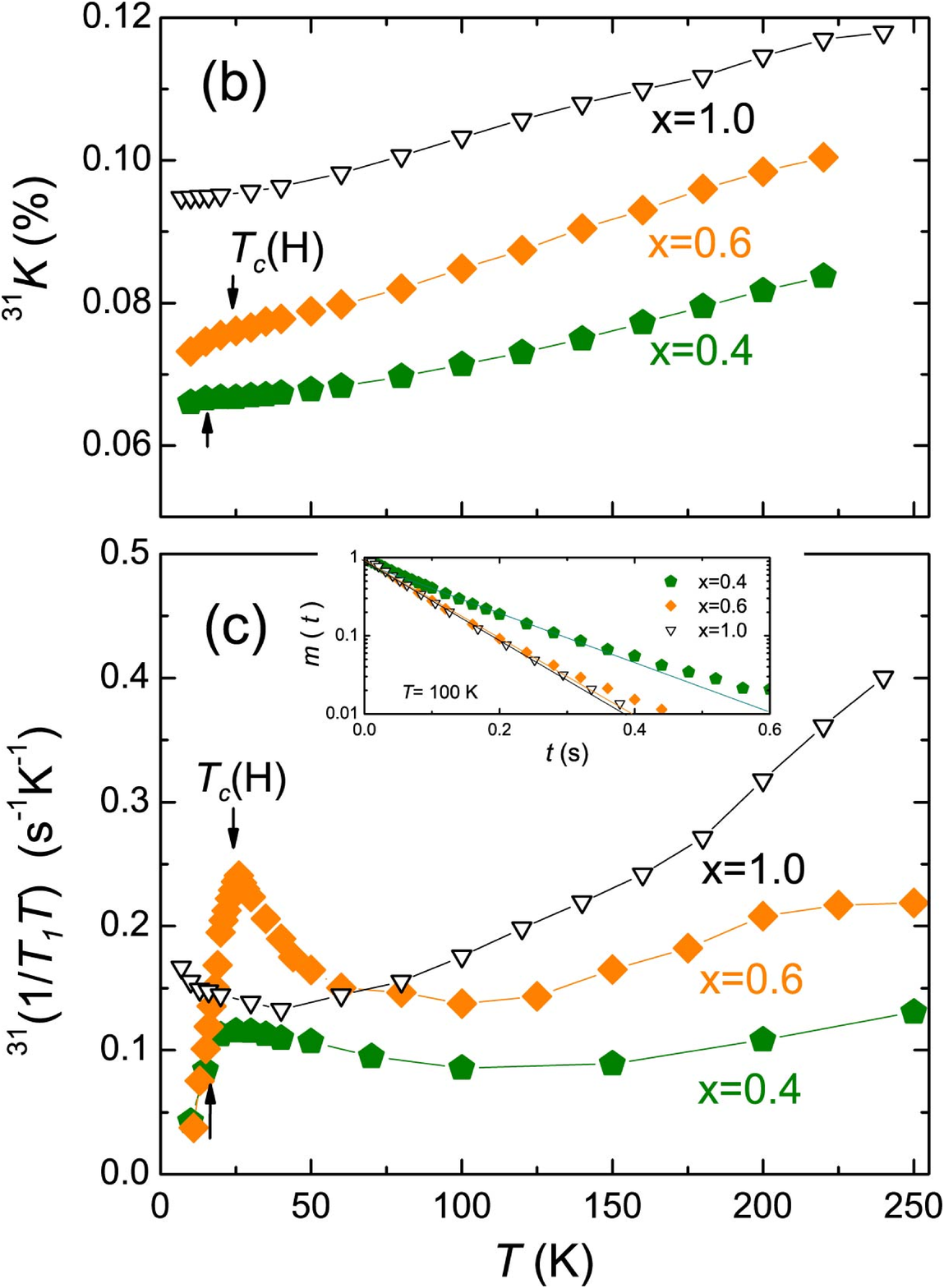}
\caption[]{(Color online) (a) $^{31}$P-NMR spectra at $T$=220 K, and $T$ dependence of the (b) Knight shift ($^{31}K$) and (c) $^{31}(1/T_1T)$ for LaFe(As$_{1-x}$P$_x$)(O$_{0.95}$F$_{0.05}$). The arrows in the inset of (a) indicate the samples used in this experiment. $T_c(H)$ indicates $T_c$ under the field $H\sim$11.95 T.  The inset in (c) shows the typical  recovery curves to determine $1/T_1$. 
} 
\label{NMR_F005}
\end{figure}

\begin{figure}[htbp]
\centering
\includegraphics[width=8cm]{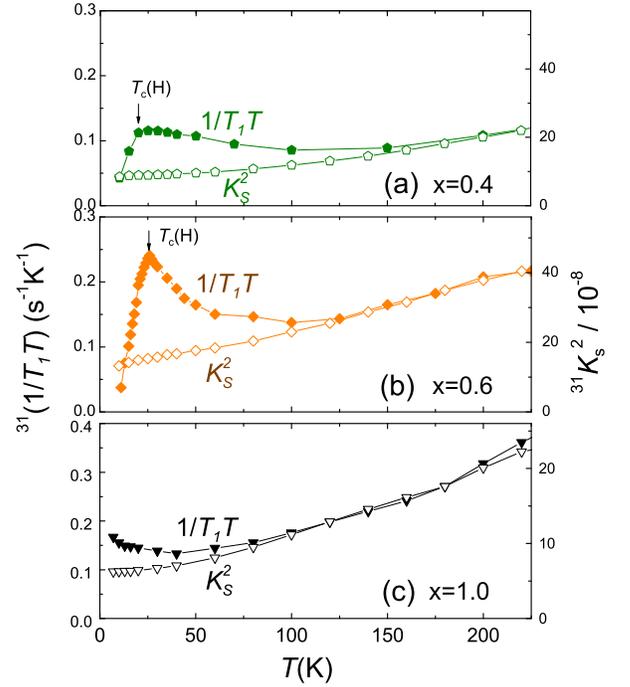}
\caption[]{(Color online) $T$ dependence of $^{31}K_s^2(=(^{31}K-^{31}K_{chem})^2)$ and $^{31}(1/T_1T)$ for (a)$x$=0.4, (b) $x$=0.6, and (c) $x$=1.0 of LaFe(As$_{1-x}$P$_x$)(O$_{0.95}$F$_{0.05}$). 
} 
\label{NMR_F005_2}
\end{figure}

\subsection{LaFe(As$_{1-x}$P$_x$)(O$_{0.95}$F$_{0.05}$)}

Next, we show the results for the underdoped compounds of LaFe(As$_{1-x}$P$_x$)(O$_{0.95}$F$_{0.05}$), i.e. for $x$=0.4 with $T_c$=19 K, $x$=0.6 with $T_c$=24 K, and $x$=1.0 with $T_c$=6.7 K. 
The $^{31}$P-NMR spectra for $x$=0.4, 0.6, and 1.0 are shown in Fig.~\ref{NMR_F005}(a). 
The FWHM is also as narrow as $\sim$73 ($\sim$135) kHz at $x$=0.4 (0.6) at the resonance frequency  $\sim$206 MHz. 
Figures~\ref{NMR_F005}(b) and \ref{NMR_F005}(c) show the $T$ dependence of the Knight shift $^{31}K$ and $^{31}(1/T_1T)$ for $x$=0.4, 0.6, and 1.0 of LaFe(As$_{1-x}$P$_x$)(O$_{0.95}$F$_{0.05}$). 
As indicated in Fig. \ref{Kchem}, $^{31}K_{chem}$ is evaluated to be $\sim$0.037\% for $x$=0.4 and 0.6 using the data at high temperatures and $\sim$0.07\% for $x$=1.0 using the data in broad $T$ range ($T>$60 K).
Figures \ref{NMR_F005_2}(a), \ref{NMR_F005_2}(b) \ref{NMR_F005_2}(c) indicate the $T$ dependence of $^{31}K_s^2$ and $^{31}(1/T_1T)$ for $x$=0.4, $x$=0.6, and $x$=1.0, respectively. 
The $^{31}(1/T_1T)$ values increase upon cooling below 100 K for $x$=0.4 and 0.6, although $^{31}K_s$ for these compounds monotonously decreases with decreasing $T$. 
In particular, $^{31}(1/T_1T)$ is more enhanced at $x$=0.6 than at $x$=0.4 and 1.0, demonstrating that the AFMSFs develop more significantly for $x$=0.6 which exhibits the higher $T_c$ than for $x$=0.4 and 1.0 with the lower $T_c$.

\subsection{AFM spin fluctuations in LaFe(As,P)(O,F) }

\begin{figure}[htbp]
\centering
\includegraphics[width=8cm]{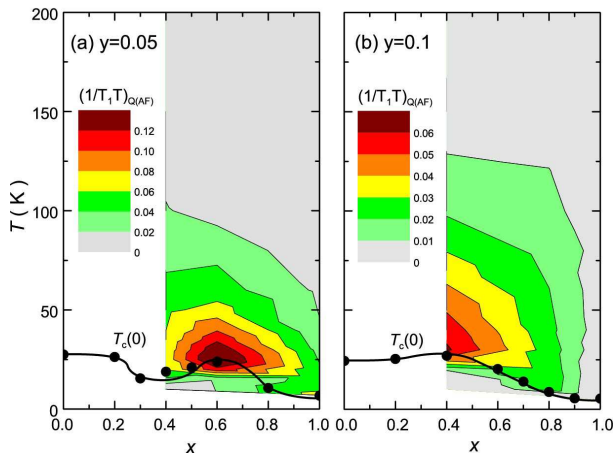}
\caption[]{(Color online) Contour plot of $^{31}(1/T_1T)_{Q(AF)}$ for  (a)$y$=0.05 and (b) $y$=0.1, indicating the development of AFM spin fluctuations at finite $Q$ wave vector is significant in the compounds, where $T_c(0)$ exhibits a peak against the variation of $x$.
Here, $T_c(0)$ represents the $T_c$ values at zero external field\cite{Saijo,Lai}. 
} 
\label{T1_AFMSF}
\end{figure}

Eventually, we remark that $T_c$ increases as AFMSFs are further enhanced  for LaFe(As$_{1-x}$P$_x$)(O$_{1-y}$F$_{y}$) compounds studied here. 
In order to deduce the development of AFM spin fluctuations for LaFe(As$_{1-x}$P$_x$)(O$_{1-y}$F$_{y}$), we assume that $^{31}(1/T_1T)$ is decomposed as,
\[
^{31}(1/T_1T)=^{31}(1/T_1T)_{Q(AF)}+^{31}(1/T_1T)_{Q-indep}, 
\]
where the former term represents the AFM spin fluctuations at finite $Q$ presumably around (0,$\pi$) and ($\pi$,0) that significantly develop upon cooling, and the latter term represents the other $q$-independent part of the background.
At high temperatures, the $T$ dependence of $^{31}(1/T_1T)$ resembles  $^{31}K_s^2(T)$, as shown in Figs. \ref{NMR_F010_2} and \ref{NMR_F005_2}, implying that $^{31}(1/T_1T)_{Q-indep}$  is predominant at high temperatures. 
Then, we can evaluate the $T$ dependence of $^{31}(1/T_1T)_{Q(AF)}$ by assuming that the $T$ dependence of $^{31}(1/T_1T)_{Q-indep}$ is identical to that of $^{31}K_s^2(T)$. 
As a result, in Figs. \ref{T1_AFMSF}(a) and \ref{T1_AFMSF}(b) we show the contour plots of $^{31}(1/T_1T)_{Q(AF)}$ for $y$=0.05 and $y$=0.1, respectively. 
These results demonstrate that the AFMSFs develop significantly for $x$=0.6 at $y$=0.05 and $x$=0.4 at $y$=0.1, where 
$T_c$ exhibits a peak against the variation of $x$.
Namely, the AFMSFs play an important role in raising $T_c$ in the LaFe(As,P)(O,F) series, although the local structure is apart from the optimum values of the Fe-based superconductors\cite{C.H.Lee,Mizuguchi}(see the inset of Fig. \ref{PhaseDiagram}).  
 
\section{Discussion}

In Fe-pnictide compounds such as Ba(Fe,Co)$_2$As$_2$~\cite{Ning}, Fe(Se,Te)~\cite{Imai,YShimizu}, BaFe$_2$(As,P)$_2$~\cite{NakaiPRL}, and Na(Fe,Co)As~\cite{Ji}, it has been well established that $T_c$ exhibits a maximum close to the AFM phase in which AFMSFs are critically enhanced. 
On the other hand, for the LaFeAs(O$_{1-y}$F$_{y}$) series, the maximum $T_c$ emerges at $y$=0.1 without any development of AFMSFs upon cooling down to $T_c$~\cite{Nakai,Grafe,Terasaki}, although AFMSFs can be observed in the vicinity of the AFM ordered phase with a lower $T_c$, i.e., in the range of $0.04<y<0.08$~\cite{Oka,Nakano,Hammarath}. 
In this context, we emphasize that the present studies of the LaFe(As,P)(O,F) compounds series provide clear evidence that the development of AFMSFs enhances $T_{\rm c}$ even if the present La1111 compounds are far away from the AFM ordered phase and optimal lattice parameters(see the inset of Fig. \ref{PhaseDiagram}). 
\begin{figure}[htbp]
\centering
\includegraphics[width=8cm]{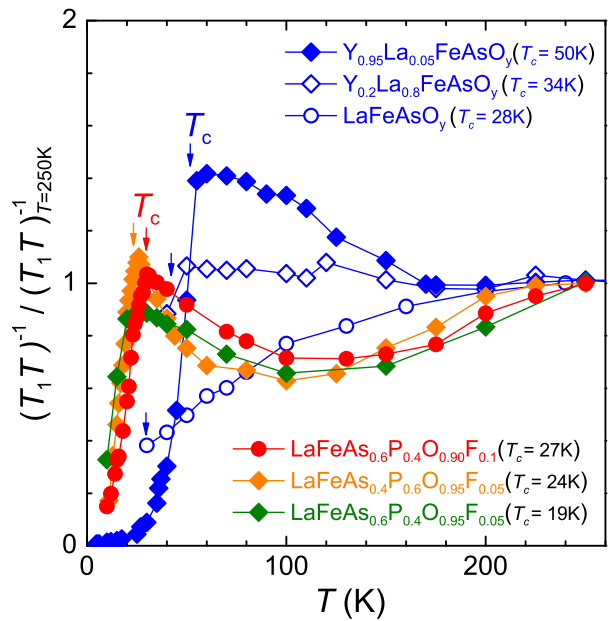}
\caption[]{(Color online) $T$ dependence of  $(T_1T)^{-1}/(T_1T)^{-1}_{T\sim250 K}$ by means of $^{31}$P-NMR for the present samples, which are compared with $^{75}$As-NMR results for Y$_{0.95}$La$_{0.05}$FeAsO$_\delta$ ($T_{\rm c}$=50 K)\cite{MukudaPRL}, Y$_{0.2}$La$_{0.8}$FeAsO$_\delta$ ($T_{\rm c}$=34 K)\cite{Yamashita_Y}, and LaFeAsO$_\delta$ ($T_{\rm c}$=28 K)\cite{MukudaNQR}.
} 
\label{comparison}
\end{figure}


Finally, we discuss a systematic comparison of the spin fluctuations among the LaFeAsO(1111)-based family, as shown in Fig.~\ref{comparison}.  Y$_{0.95}$La$_{0.05}$1111 with $T_c$=50 K~\cite{MukudaPRL} possessing near-optimal structural parameters in the FeAs block ($h_{pn}\sim1.44$\AA) is characterized by three hole Fermi surfaces; two of them are located at $\Gamma$(0,0) and the other is at $\Gamma^{\prime}$($\pi$,$\pi$), and two electron Fermi surfaces at $M$[(0,$\pi$)($\pi$,0)] in the unfolded Fermi surface regime~\cite{Kuroki2,Usui}. 
The appearance of $\Gamma^{\prime}$ at $E_F$ causes the Fermi surface nesting condition to be better in Y$_{0.95}$La$_{0.05}$1111($T_{\rm c}$=50 K) than the other compounds. This results in the enhancement of AFMSFs for Y$_{0.95}$La$_{0.05}$1111, that is, $T_c$ increases from 28 K in La1111, to 34 K in La$_{0.8}$Y$_{0.2}$1111 up to 50 K in Y$_{0.95}$La$_{0.05}$1111~\cite{MukudaPRL,Yamashita_Y,MukudaNQR}. 
According to the spin-fluctuation mediated SC mechanism,  the large Fermi surface multiplicity in $Ln$1111 in addition to the presence of AFMSFs is an another crucial factor for enhancing $T_c$, which is optimized when the FeAs$_{4}$ tetrahedron is close to a regular one realized in $Ln$1111~\cite{Usui}. 
It is noteworthy that the $T$ dependence of $1/T_1T$ of (Y$_{0.95}$La$_{0.05}$)1111 is saturated below 100 K.  
A similar saturation and/or broad maximum in $1/T_1T$ was observed for Ba$_{0.6}$K$_{0.4}$Fe$_2$As$_2$ ($T_c$= 38 K)~\cite{Yashima,Hirano}, Ca$_4$(Mg,Ti)$_3$Fe$_2$As$_2$O$_{8-y}$ ($T_c$= 47 K)\cite{Tomita}, and Sr$_4$(Mg$_{0.3}$Ti$_{0.7}$)$_2$O$_6$Fe$_2$As$_2$ ($T_c$=34K)~\cite{Yamamoto}, which are characterized by the lattice parameters of the FeAs block being close to the values of the regular tetrahedron.
This is in contrast to the $T$ dependence of $1/T_1T$ in  LaFe(As$_{1-x}$P$_x$)(O$_{1-y}$F$_{y}$) compounds that continues to increase down to $T_c$ as seen in Fig.~\ref{comparison}. 
Likewise, $T_c$ for the Fe-pnictides that reveal a significant enhancement of AFMSFs towards $T_c$ is nearly limited in the compounds within the range from $T_c\sim$10 K to $T_c\sim$30 K.
These results suggest that AFMSFs are not always  a unique factor to attain $T_c$=55 K in the Fe-based compounds.
In this context, the optimized electronic states for the occurrence of SC in Fe-pnictides is realized for the regular FeAs$_4$ tetrahedron in which the multiorbital fluctuations may play some roles for the onset of SC\cite{Kontani}, since  the spin and orbital degrees of freedom can be intimately coupled to each other.

\section{Conclusion}

In conclusion,  systematic $^{31}$P-NMR studies of LaFe(As$_{1-x}$P$_x$)(O$_{1-y}$F$_y$) have revealed that the antiferromagnetic spin fluctuations  at low energies cause a peak at $T_c$=27 K and at $T_c$=24 K for $y$=0.1 and 0.05, respectively. 
The result indicates that the AFMSFs are responsible for the $T_c$ increase in LaFe(As$_{1-x}$P$_x$)(O$_{1-y}$F$_y$) as a primary mediator of the Cooper pairing. 
We highlight that the present studies of the LaFe(As,P)(O,F) series compounds  provide clear evidence that the development of AFMSFs enhances $T_{\rm c}$ even if the present La1111 compounds are far from the AFM ordered phase and optimal lattice parameters. 
In the $T_c$=50 K class of Fe-pnictides, however, it should be noted that the AFMSFs do not critically develop down to $T_c$, instead, they seem to be saturated. We propose that $T_c$ of Fe-pnictides exceeding 50 K is maximized under an intimate collaboration of the AFMSFs and other factors originating from the optimization of the local structure.

\section*{Acknowledgement}

{\footnotesize 
We thank K. Kuroki and K. Suzuki for fruitful discussion and comments. 
}

\end{document}